# Terahertz Response of Field-Effect Transistors in Saturation Regime

T. A. Elkhatib,[1,a] V. Yu. Kachorovskii,[2] W. J. Stillman,[1] S. Rumyantsev,[1,2] X.-C. Zhang,[1] and M. S. Shur[1]

[1]*Rensselaer polytechnic Institute, 110, 8th street, Troy, NY, 12180, USA.*

[2]*A.F. Ioffe Physical-Technical Institute, 26 Polytechnicheskaya Street, St. Petersburg, 194021, Russia.*

We report on the broadband THz response of InGaAs/GaAs HEMTs operating at 1.63 THz and room temperature deep in the saturation regime. We demonstrate that responses show linear increase with drain-to-source voltage (or drain bias current) and reach very high values up to 170V/W. We also develop a phenomenological theory valid both in the ohmic and in the saturation regimes.

---

[a] Author to whom correspondence should be addressed: Electronic mail: elkhat@rpi.edu

Article

The exploration of plasma phenomena in field-effect transistors (FETs) has begun long time ago starting from the theoretical prediction[1] that such waves have linear dispersion, $\omega = sk$, with the velocity $s \simeq \sqrt{eU_g/m}$ controlled by the gate voltage swing $U_g$. This work was followed by observation of infrared asbsorption[2] and weak infrared emission[3] related to excitation of plasma oscillations in silicon inversion layers. In recent decade, the interest to plasma phenomena increased due to realization that high typical values of $s$ ($\approx 10^8 cm/c$) are promising for utilization of plasma oscillations in emission and detection of THz radiation.[4,5]

The plasma wave approach to THz electronics[6-27] has led to resonant and broadband (non-resonant) detection or emission of sub-THz and THz radiation by nanoscale FETs. Though the plasma wave THz emitters are still rather weak, the THz detectors already show good performance: they are tunable[6-27] by varying the gate voltage and the drain current, with fast response time[22] and demonstrate relatively low noise equivalent power (NEP) up to room temperature[18]. Hence, they are promising candidates for terahertz systems applications.

As shown in Refs. [6] and [17], the efficiency of the broadband detection can be significantly improved by applying drain current $j$ to drive FETs into the saturation regime. The theory[22] demonstrates that the response essentially increases when $j$ approaches the saturation current $j_{sat}$; (knee current on *I-V* curve). However, THz response of FETs operating in the deep saturation regime, $j>j_{sat}$, was neither investigated experimentally nor theoretically until this work.

Here we show that for $j > j_{sat}$ a further dramatic growth of broadband response can be achieved. We report that in this region response continues to grow linearly with dc drain voltage (or dc drain current) and it may reach very high values (up to 170V/W at 1.63

THz). We also develop a phenomenological theory which yields an excellent fit of experimental data. The key point of this phenomenological description is that all important information needed for the THz response calculation is encoded in the stationary current-voltage transistor characteristics. We will show how to extract this information in order to calculate the response. This approach allows one to find response in a wide range of parameters without knowing detailed information about number of nontrivial processes in transistor channel (such as the velocity saturation, heating of electron gas and the lattice etc.) coming into play at $j > j_{sat}$.

We start with the discussion of our theoretical approach. The situation relevant to our experiment is $\omega_o \tau \ll 1$ (non-resonant detection), where $\omega_o$ is the plasma fundamental frequency and $\tau$ is the momentum relaxation time. In this case, the variation of alternating component of the voltage[5,17] decreases exponentially along the channel on the characteristic spatial scale $L_\circ = \sqrt{\mu U_g / \omega}$, where $\mu$ is the electron mobility, and $\omega$ is the frequency of the incident radiation. For $L_o \ll L$, the alternating components of electric potential is concentrated near the edge of the gated region on the source side[5,17] (in our experiment $\omega_o \tau \approx 0.1$-$0.2$, $L_o \approx 0.05$-$0.1 \mu m$, and $L = 0.5 \mu m$).

As shown in Refs. 6 and 24, THz radiation induced voltage drops between the gate and source and between drain and gate. However, our experimental results are consistent with the situation when the source-to-gate THz signal is dominant. In the saturation regime, the gated channel could be divided into two regions: the region of length $L^*$ near the source, where the electron velocity is not saturated and the velocity saturation region with the length of $L - L^*$. (We assume that $L_o < L^*$). In the unsaturated region, the electric field is smaller than the velocity saturation field, and the theory developed in Ref. 17 is still valid. The potential in this region is given by $U = U_\circ(x) + (U_a e^{-i\omega t - qx} + U_a e^{i\omega t - q^* x})/2$, where $U_a$ is the gate voltage oscillation amplitude due to the incident THz radiation, and $q = (1-i)/\sqrt{2}L_\circ$.

For transistors operating above threshold, the stationary part of the potential is given by (see Eq. (20) of Ref. 17): $U_\circ(x) = \sqrt{U_g^2 + U_a^2[1 - \exp(-\sqrt{2}x/L_\circ)]/2 - 2jx/\mu C}$. At the boundary of this region ($x \approx L^*$), we have $\exp(-\sqrt{2}x/L_\circ) \ll 1$; so that $U_\circ(x) = \sqrt{U_g^2 + U_a^2/2 - 2jx/\mu C}$. The stationary potential in the rest of the gated region ($L^* < x < L$) is not disturbed by incoming radiation and coincides with the potential without radiation but with the gate voltage swing $U_g$ replaced by the effective swing $U_g' = \sqrt{U_g^2 + U_a^2/2} \approx U_g + U_a^2/4U_g$. As a consequence, the voltage drop across the channel $V$ can be written as a difference of $U_g$ and the gate-to-drain potential $U(j, U_g')$, which is a function of the current and the effective gate voltage swing: $V = U_g - U(j, U_g')$. In the absence of radiation, this equation becomes $V_\circ = U_g - U(j, U_g)$, and the transistor broadband terahertz response which is defined as: $\delta V = V_\circ - V = U(j, U_g') - U(j, U_g) \approx (U_a^2/4U_g)(\partial U/\partial U_g)$, can be written as:

$$\delta V = \frac{U_a^2}{4U_g}\left(1 - \frac{\partial V}{\partial U_g}\bigg|_{j=const}\right) \quad (1)$$

(Please notice that $\partial V/\partial U_g |_{j=const} < 0$, since the same current is reached at smaller $V$ when $U_g$ is increased). Eq. (1) links the response to the steady-state voltage-current characteristics. It is valid for both linear and saturation regimes and does not rely upon any simplified models. It is only based on the fact that the radiation-induced excitation of the electron potential is concentrated near the gate edge from the source side. Let us now demonstrate how this equation works for two limiting cases: (1) $j < j_{sat}$ and (2) $j > j_{sat}$. For $j < j_{sat}$ one can roughly describe the response by the simplified Shockley model for transistor operating in the linear regime. Hence, $V = U_g - \sqrt{U_g^2 - 2jL/\mu C}$, $\partial V/\partial U_g = 1 - U_g(U_g^2 - 2jL/\mu C)^{-1/2}$ and we restore Eq. (40) from Ref. 17:

$$\delta V = \frac{U_a^2}{4U_g \sqrt{1 - j/j_{sat}}} \quad (2)$$

where $j_{sat} = \mu C U_g^2 / 2L$ is the pinch-off current. Considering the other case of the saturation regime, one can interpolate the measured current-voltage curves by the simple equation: $j = j_{sat}(1 + \lambda V)$ such that $\partial V / \partial U_g = -2j/\lambda j_{sat} U_g$, and we obtain the linear THz response:

$$\delta V = \frac{U_a^2}{4U_g}\left(1 + \frac{2}{\lambda U_g}\frac{j}{j_{sat}}\right) \quad (3)$$

where $\lambda$ is the channel length modulation parameter. One can also generalize Eq. (5) by taking into account the finite value of the voltmeter resistance $R^*$ (see Ref. 20), and by writing $U_g = U_{gs} - V_{th}$ (here $U_{gs}$ is the gate-to-source voltage), where the threshold voltage $V_{th}(V) = V_{th}(0) + \kappa V$ is assumed to depend on $V$, We obtain

$$\delta V = \frac{U_a^2}{4U_g}\left(1 - [1+\kappa]\frac{\partial V}{\partial U_g}\right)\frac{R^*}{R^* + \partial V/\partial j} \quad (4)$$

Next, we discuss our experimental data. We used our InGaAs/GaAs HEMTs with the gate length of 0.5$\mu m$ fabricated by TriQuint Semiconductor. Figure 1 shows the current-voltage characteristics of our HEMTs with all important device parameters extracted in Ref. 27. An optically pumped far-infrared gas laser (Coherent SIFIR-50) was used as the source of 1.63 THz radiation for our response measurements. The laser beam (polarized parallel to the transistor channel) was modulated by an optical chopper with a fixed frequency of 100 Hz, and it was focused with normal incidence on the transistor surface. The induced dc drain voltage $\delta V$ superimposed on the drain-to-source voltage *V* was then measured using the lock-in amplifier SR-830.

The measured THz responses as function of the applied drain-bias voltages for different gate-bias voltages are shown in Fig. 2. As seen, the response increases linearly in the deep saturation regime after it was significantly increased at the transition from the linear regime into the saturation regime. The calculated THz responses based on Eq. (4) closely match the measured response, while the derivative terms in our model were obtained numerically from the measured current voltages characteristics of the fabricated HEMTs. The results shown in Fig. 2 can be also presented as function of the drain-bias current (see Fig. 3), where calculated response from Eqs. (2) and (3) are also shown for comparison. It is clear that our latest model describes well the response in all operating regimes and reproduces the smooth transition of the response between the linear and saturation regimes.

The linear response of FETs in the saturation regime offers many advantages for THz detection and imaging applications: THz responsivity can be increased by 2-4 times when transistors operate deeper in the saturation regime; an additional increase in responsivity may be gained when many transistors are connected in series with the same drain-bias current and the same gate-to-source voltage. As shown in Ref. 20, the response is proportional to the number of series connected transistors. In addition, the saturation regime can be used for imaging the incident THz radiation at sub-wavelength resolution as shown in Refs. 25 and 26.

On the other hand, the main disadvantage of THz detection in the saturation regime is the higher flicker noise at high drain current values; the measured noise equivalent power was on the order of $10^{-7}$ W/Hz$^{0.5}$. However, the signal-to noise-ratio (SNR) can be improved by a factor of $\sqrt{N}$ for the series connection of many transistors in the saturation regime[27], where *N* is the number of connected transistors. In addition, the electronic modulation of incident THz radiation above the 1/f corner frequency can also enhance the NEP[19].

In conclusion, we have reported on a linear growth of broadband THz response in a FET operating in the deep saturation regime. We have reached a very high value of the response (up to 170V/W). We have also developed a phenomenological theory that shows an excellent agreement with experimental data. This detection regime has boosted the performance of plasma wave detectors for applications in THz systems, including imaging systems with sub-wavelength resolution.

The work at RPI was supported by subcontract from AFOSR (Program Manager Dr. Gernot Pomrenke). The work at A. F. Ioffe was supported by RFBR, Programs of the RAS, and Rosnauka Grant 02.740.11.5072.

Figures

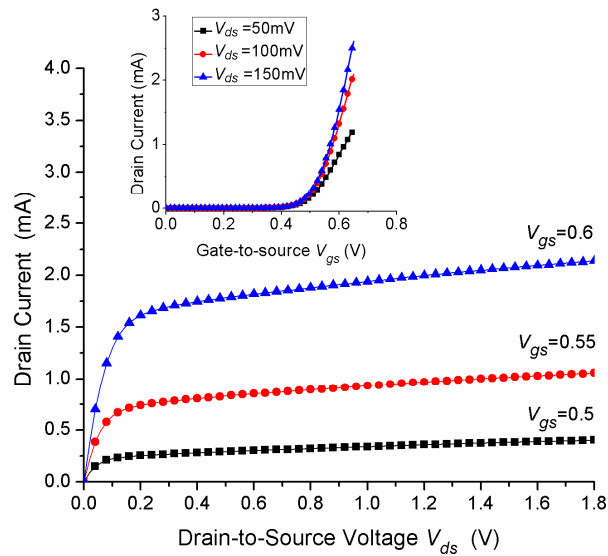

FIG. 1 (Color online) Measured current-voltage characteristics ($I_d$ -$V_{ds}$) of the InGaAs/GaAs HEMT at different gate-to-source voltages. Inset shows the measured ($I_d$ -$V_{gs}$) characteristics for different drain-to-source voltages.

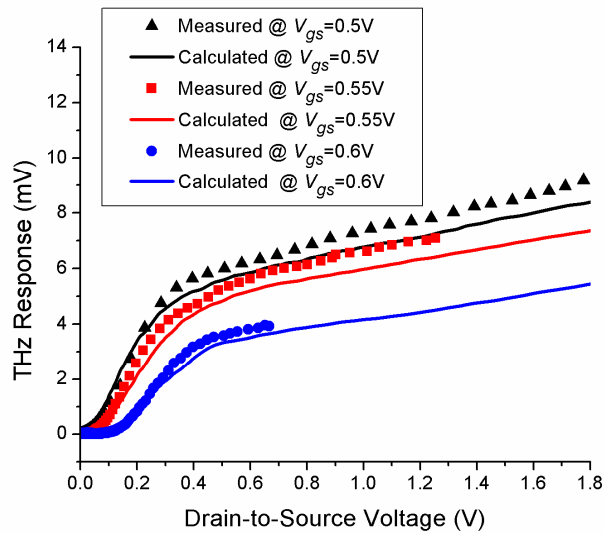

FIG. 2 (Color online) Measured and calculated (Eq. 4) broadband THz Response at 1.63 THz as a function of the drain-to-source voltage for different gate-bias voltages. The responses start to grow linearly when transistor is driven deeper in the saturation regime**.** The transistor threshold voltage is 0.48V and the load resistor was 10Kohms.

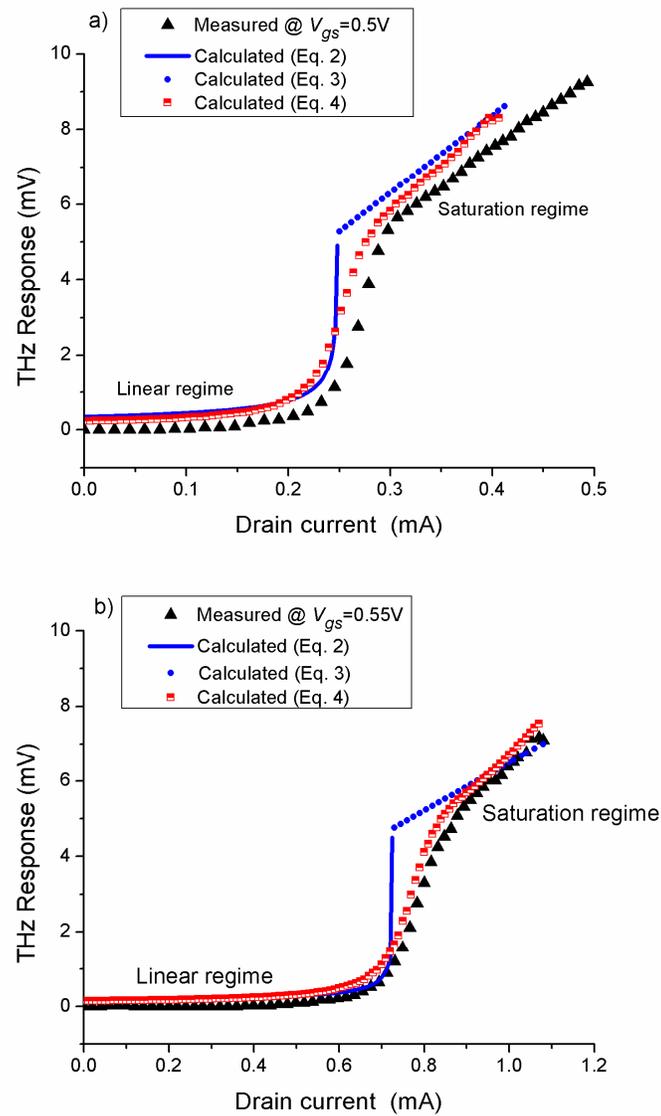

FIG. 3 (Color online) Measured and calculated THz response as a function of the drain current for gate-biases: (a) 0.5V and (b) 0.55V. The linear responsivity in the saturation regime appears clearly on the proper linear scale. Eq. 4 matches the responses in all regimes very well.